\documentclass[12pt,preprint]{aastex}

\shorttitle{Radio Emission from Disks}
\shortauthors{Pascucci et al.}

\begin{document}


\title{Free-free Emission and Radio Recombination Lines from Photoevaporating Disks}


\author{I. Pascucci}
\affil{Lunar and Planetary Laboratory, The University of Arizona, Tucson, AZ 85721, USA}
\email{pascucci@lpl.arizona.edu}

\and

\author{U. Gorti\altaffilmark{1} and D. Hollenbach}
\affil{SETI Institute, 189 Bernardo Ave., Mountain View, CA 94043, USA}
\altaffiltext{1}{NASA Ames Research Center, Moffett Field, CA 94035, USA}

\begin{abstract}
Recent infrared observations have demonstrated that photoevaporation driven by high-energy photons from the central star contributes to the dispersal of protoplanetary disks. Here, we show that photoevaporative winds should produce a detectable free-free continuum emission given the range of stellar ionizing photons and X-ray luminosities inferred for young sun-like stars. We point out that VLA observations of the nearby disk around TW~Hya might have already detected this emission at centimeter wavelengths and calculate the wind electron density and mass flow rate.
We also estimate the intensities of H radio recombination lines tracing the wind and discuss which ones could be detected with current instrumentation. The detection and profiles of these recombination lines would unambiguously prove our inference of free-free emission from photoevaporating disks like TW~Hya. In addition, radio/millimeter data can help constraining wind parameters such as temperature and electron density that are fundamental in measuring mass flow rates. 
\end{abstract}





\section{Introduction}
Photoevaporation driven by high-energy photons from the central star has been long recognized by theorists as a plausible mechanism to speed-up the clearing of protoplanetary disks even around low-mass sun-like stars (e.g., \citealt{h00}). Recently, we identified a robust diagnostic for photoevaporation in the 
forbidden [Ne~II] emission line at 12.81\,\micron{} \citep{ps09}. Relatively narrow ($\sim$20\,km/s) and slightly blueshifted (a few km/s) [Ne~II] line profiles have been detected toward  several evolved disks, including those having a gap in their dust distribution, often called transitional disks \citep{ps09,ps11,sacco12}. Such profiles unambiguously point to unbound gas in a wind. Other wind tracers have been proposed and are being investigated, e.g. the [O~I] line at 6300\,\AA{} \citep{eo10} and the CO rovibrational band at 4.67\,\micron{} \citep{pont11}. Because mass flow rates are very sensitive to the density and temperature of the wind, it is necessary to identify additional wind diagnostics to pin down these parameters.

Here, we show that observations at millimeter and radio wavelengths can provide such diagnostics. 
The photoevaporative wind may be fully (EUV case)\footnote{EUV; $13.6$\,eV $ < h \nu \lesssim 100$\,eV} or partially (X-ray case) ionized and deflections of electrons in the wind by protons should result in continuum free-free emission (Sect.~\ref{sect:plausible}). We point out that VLA observations of the nearby disk around TW~Hya might have already detected the free-free emission from its photoevaporative wind (Sect.~\ref{sect:excess}). This emission should be accompanied by H radio recombination lines. We calculate the fluxes of a representative sample of transitions and show that they can be detected with current instruments (Sect.~\ref{sect:recombination}). We conclude by discussing the broader implications of this study (Sect.~\ref{sect:conclusions}).

\section{Free-Free Emission from Ionized Winds}\label{sect:plausible}
For protoplanetary disks, H$^+$ is the dominant ion in the surface layer ionized by stellar EUV and X-ray photons. In this case the thermal free-free volume emissivity can be fully characterized by a handful of parameters (see, e.g. \citealt{pad00}):
\begin{equation}\label{eq:freeemiss}
\epsilon_\nu = 6.8 \times 10^{-38} n_e^2 T^{-1/2} e^{-h \nu/ k_B T} g_{ff}  \,  \, \,
[erg \, cm^{-3} \, s^{-1} \, Hz^{-1}]
\end{equation} 
where $n_e$ is the electron density, $T$ is the gas temperature, and $g_{ff}$ is the velocity-averaged Gaunt factor. This factor depends only on $T$ and the frequency $\nu$  of the observation and can be computed analytically. Converting the emissivity into luminosity ($L_\nu$) requires integrating $\epsilon_\nu dV$, where $V$ is the volume of the emitting gas. Assuming temperature is constant, $L_\nu$ is proportional to the integral of $n_e^2 dV=$EM$_V$, the volume emission measure. 

Assume that the disk surface absorbs a fraction $f$ ($\sim 0.7$ for EUV photons, see \citealt{hg09}) of the stellar EUV photon luminosity $\Phi_{EUV}$.
 In steady state, $f \, \Phi_{\rm EUV}$ is equal to the recombination rate of electrons and protons in the ionized gas, and the latter is proportional to EM$_V$.
Therefore, the free-free luminosity is linearly dependent on the EUV photon luminosity.
Similarly, the absorbed X-rays create a hydrogen ionization rate balanced by recombinations, so that the free-free emission caused by Xray ionizations is linearly proportional to the locally incident X-ray photon luminosity.   Here we use the fact that, including secondary ionizations, X-ray ionization rates (in s$^{-1}$) are equal to the incident X-ray luminosity divided by 40\,eV, the mean energy required for a single ionization \citep{glassgold04}. Because X-rays penetrate deeper in the disk than EUV photons, the fraction $f$ of photons absorbed by the disk is lower. Detailed modeling suggests a value of 0.5 for a $\sim$5,000\,K gas \citep{gorti11}.
Assuming the EUV ionized layer has T=10,000\,K and the X-ray layer has T= 5,000\,K, the free-free flux density at 3.5\,cm can be written as:   
\begin{eqnarray}\label{eq:freeflux}
F_{3.5cm}=2.9\times10^{-39}\, \left( \frac{51}{d} \right)^2\,\Phi_{\rm EUV} \, \, \, [\mu Jy]\\
F_{3.5cm}=2.4\times10^{-29}\, \left( \frac{51}{d} \right)^2\, L_{\rm X} \, \, \, [\mu Jy]
\end{eqnarray}
here $d$ is the source distance in pc (where 51\,pc is the distance of TW~Hya, \citealt{mamajek05}),
$\Phi_{\rm EUV}$ is measured in photons per second, and $L_{\rm X}$ in erg/s. We note that the bulk of
the X-ray-heated gas is typically at $\sim$1,000-2,000\,K. However, the free-free flux density decreases with
gas temperature such that $F_{3.5cm}$(1,000\,K)$<0.5\times \, F_{3.5cm}$(5,000\,K).

Direct measurements of $\Phi_{\rm EUV}$ from young stars are not available because EUV photons are easily absorbed by H in the interstellar medium. 
\citet{alexander05} used emission lines at FUV wavelengths to derive order-of-magnitude estimates of $\Phi_{\rm EUV}$ for five young solar mass stars of $10^{41}$--$10^{44}$\,s$^{-1}$.
Based on this large range, we can expect a free-free 3.5\,cm flux from the fully ionized layer ranging from 290\,$\mu$Jy to 0.3\,Jy at the distance of TW~Hya. Stellar X-ray luminosities have been measured in several star-forming regions. The XMM survey of Taurus finds $L_{\rm X}$ values from $10^{29}$ to $10^{31}$\,erg/s for solar-mass stars \citep{guedel07} which convert into a free-free flux from 2.4 to 240\,$\mu$Jy at 51\,pc. Given that the extended VLA (EVLA) can reach sensitivities of a few $\mu$Jy at 3.5\,cm in a few hours of integration, this simple calculation illustrates that free-free emission from a fully or partially ionized disk surface can be detected with current instrumentation out to the distance of nearby star-forming regions like Taurus.

\section{The case of TW~Hya}\label{sect:excess}
TW~Hya is a nearby, relatively old star surrounded by a transitional disk (e.g, \citealt{calvet02}). The exact structure and extension of the dust gap is debated (\citealt{akeson11} for a discussion) as well as the mechanisms responsible for the inner clearing (e.g., \citealt{gorti11}).

Recently, we found that the disk of TW~Hya is dispersing gas from its surface via photoevaporation driven by high-energy photons from the central star \citep{ps09}. The photoevaporative wind, as traced by the [Ne~II] line at 12.81\,\micron{}, originates beyond the radius that marks the transition between the thin and thick dust disk, regardless of whether that occurs at 1 or 4\,AU, and extends out to about 10\,AU \citep{ps11}. 

\begin{figure}
\includegraphics[scale=0.8]{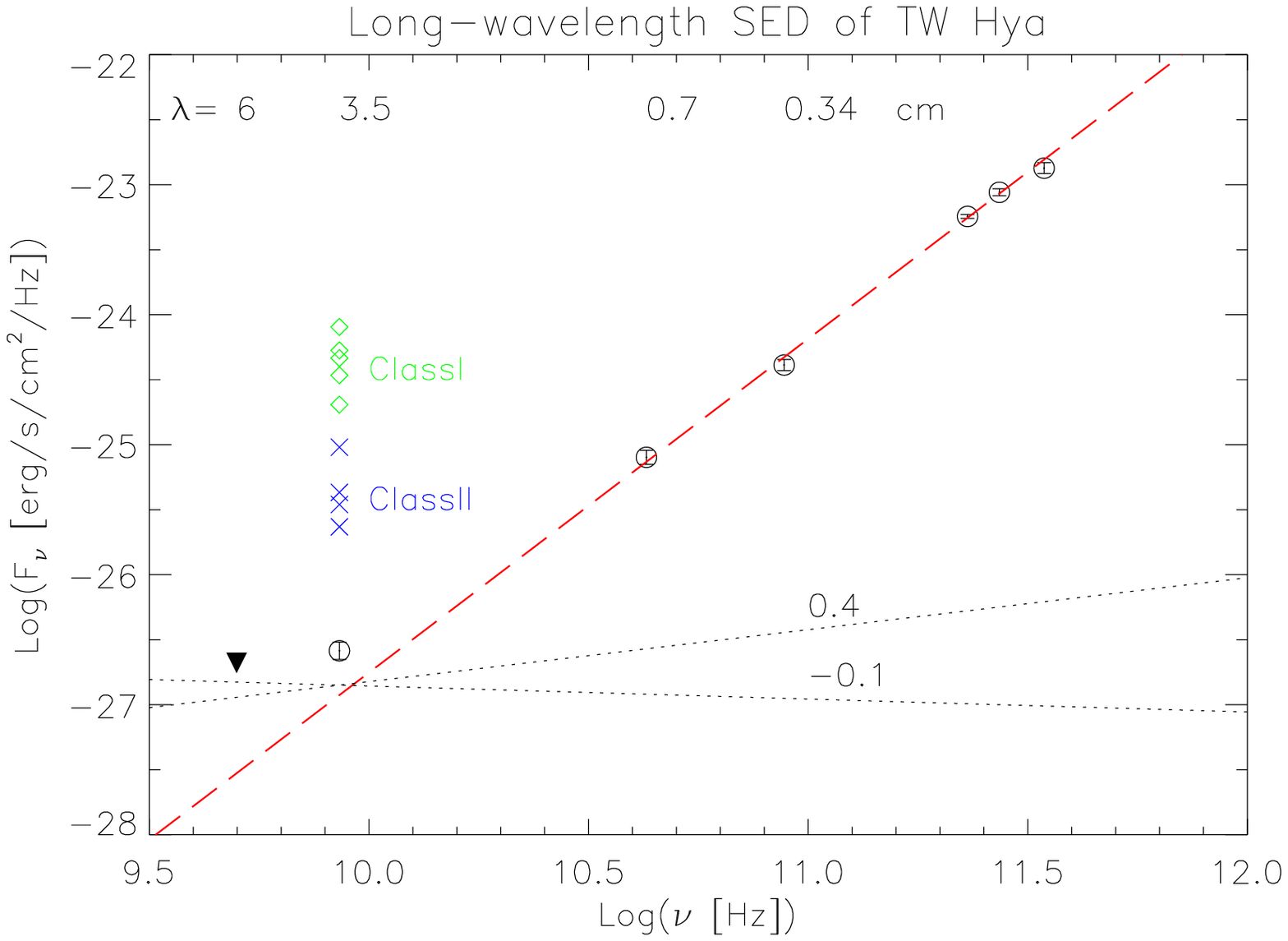}
\caption{
SED of the TW~Hya disk from 0.87\,mm out to 6\,cm. Observed fluxes (empty circles) and 3$\sigma$ upper limits (downward triangle) are from Wilner et al.~(2005) and Andrews et al.~(2012). The dashed line is a linear fit to the millimeter fluxes between 0.87 and 3.4\,mm. Note that this fit fully accounts for the 7\,mm flux but under-predicts the 3.5\,cm flux by a factor of 2.2. Dotted lines are free-free emission relations expected for optically thin ($\alpha=$-0.1) and thick ($\alpha=$0.4) gas passing through the observed minus dust emission at 3.5\,cm. We also plot literature 3.5\,cm fluxes from Class~I (Anglada et al. 1998) and Class~II/classical disks (Rodmann et al.~2006) scaled at the distance of TW~Hya.\label{fig:sed}
}
\end{figure}

Having evidence for a fully or partially ionized disk layer and given its proximity, TW~Hya is the ideal source to search for the type of free-free emission discussed in Sect.~\ref{sect:plausible}. Fig.~\ref{fig:sed} shows the long-wavelength portion of the source spectral energy distribution (SED) covering from 0.87\,mm out to 6\,cm (data are from \citealt{wilner05}, \citealt{andrews12}, and references therein). Emission from most disk grains at these long wavelengths is optically thin hence the flux depends on the wavelength as  $F_\nu \propto \nu^{(2+\beta)}$, where $\beta$ is the wavelength dependence of the dust opacity. 
In the Log-Log plot shown in Fig.~\ref{fig:sed} the best fit to the millimeter data (from 0.87 to 3.4\,mm) is for a SED slope of 2.57$\pm$0.06, implying a $\beta$ of 0.6. This $\beta$ is typical of classical and more evolved disks in the Taurus-Auriga star-forming region \citep{ricci10}. Our fit fully accounts for the observed 7\,mm flux but it is lower than the 3.5\,cm flux by more than a factor of 2. The extra emission at 3.5\,cm amounts to 140$\pm$40$\mu$Jy. 

A fully ionized wind by $4-6\times10^{40}$ EUV photons per second or a wind partially ionized by a star with $L_{\rm X}$ of $4-8\times10^{30}$\,erg/s can alone account for this excess emission (see eqs.~2 and 3). The X-ray luminosity of TW~Hya is measured to be $\sim 1.5 \times 10^{30}$\,erg/s \citep{brickhouse10} meaning that X-rays contribute to only $\sim 35 \mu$Jy of the excess flux, even in the optimistic assumption that all the X-ray heated gas is at 5,000\,K. This small flux is within the $1\sigma$ flux uncertainty, hence most of the measured excess emission at 3.5\,cm must arise from the fully ionized EUV-layer in this disk. 

The $\Phi_{\rm EUV}$ derived from eq.~2 converts into a  mass loss rate 
$\dot{M}_{\rm wind}$ of $2-3\times10^{-10}$\,M$_\sun$/yr \citep{hollenbach94}. 
The mass accretion rate of TW~Hya is time variable and literature values span a large range, 
from close to the inferred  $\dot{M}_{\rm wind}$ ($5\times10^{-10}$\,M$_\sun$/yr, \citealt{muzerolle00}) 
to more than 10 times higher \citep{alencar02,dupree12}. 
This range suggests that EUV-driven photoevaporation does not dominate yet over viscous accretion. 
We note that indirect measurements of $\Phi_{\rm EUV}$ for TW~Hya, including from the [Ne~II] line, 
range from $2-5\times10^{41}$\,s$^{-1}$ with a large uncertainty of about a factor of 5 \citep{herczeg07,ps09}. 
Our estimates from the free-free continuum emission are consistent with these values within the reported 
errorbars. 

The wind electron density $n_e$ can be also estimated from the free-free luminosity and $EM_V$. \citet{hg09} found that for disks EUV-illuminated by the central star most of $EM_V$ comes from
regions near the "gravitational radius" $r_g$, where the hydrogen thermal speed is equal to the escape speed from the star gravitational field.
This is indeed what we see in the [Ne~II] line, whose critical density is above the wind density: $r_g$ is 6.2\,AU for TW~Hya and most of the [Ne~II] emission arises within a radius of 10\,AU \citep{ps11}.  Additionally at the gravitational radius we have a vertical extent $z_{HII} \sim r_g$ \citep{hg09}. Based on these considerations, we adopt $r_{HII}=$10\,AU and $z_{HII}=$5\,AU and find $n_e \sim 10^5$\,cm$^{-3}$.




\section{Radio Continuum Emission from Other Mechanisms}\label{sect:Others}
We briefly examine five additional mechanisms that could produce excess emission at centimeter wavelengths and discuss what observations are needed to discriminate among them.

{\it Collimated ionized outflows/jets} from Class~I and II sources present flat or positive spectral indexes ($\alpha \ge -0.1$ with $F_\nu \propto \nu^\alpha$) at radio wavelengths pointing to free-free emission (e.g., \citealt{anglada98,rodmann06} and Fig.~\ref{fig:sed}). 
Hence, searches for free-free emission from photoevaporative winds should be carried out in evolved systems like TW~Hya that have no jets/outflows detected in molecular, atomic, and/or ionic tracers \citep{alencar02,ps11}. Most transitional disks belong to this category. We can estimate the jet contribution to the free-free emission by
using the empirical relation between momentum rate in the molecular outflow and the radio continuum luminosity
(eq. 3 from \citealt{anglada98}). For TW~Hya, even the largest measured mass accretion rate and ratio between
outflow and accretion \citep{hartigan95,white} would produce a 3.5\,cm free-free flux of only
$\sim$30\,$\mu$Jy for a typical jet velocity of 200\,km/s, $\sim$20\% of the measured excess flux. 
Similarly, any free-free contribution from {\it accretion shocks near the forming star} is reduced for evolved
disks. In the case of TW~Hya it should amount to less than a few $\mu$Jy at 3.5\,cm based on the calculations
of \citet{neufeld96}.
 

Another source of radio emission in young stars is {\it non-thermal} emission originating in magnetic fields, also known as {\it (gyro)synchrotron radiation}. A few protostars  (Class0/I)  have been reported to have this type of emission based on spectral indexes $\alpha < -0.1$ and about an order of magnitude variability, presumably due to magnetically induced flares (e.g., \citealt{forbrich10}). The easiest way to discriminate between free-free and gyro-synchrotron radiation is by obtaining radio observations at multiple wavelengths and computing the radio spectral index. In the case of TW~Hya, where the only radio detection is at 3.5\,cm, it is time monitoring that can rule out gyro-synchrotron emission \citep{wilner05}.

Finally, both very large (mm-size) and very small (nm-size) grains can enhance the centimeter flux. In the
case of TW~Hya, \citet{wilner05} showed that a population of 5-7\,mm sized-grains, containing 99.9\% of the
dust mass, can match the 3.5\,cm excess emission. The remaining 0.1\% of the dust mass is in grains of sizes
between 0.005 and  1\,\micron{} in their model suggesting a strictly bimodal dust distribution. At the other
end of the grain size spectrum, \citet{rafikov06} showed that nm-sized grains spinning at thermal rates can
produce detectable {\it electric dipole emission} at $\lambda \ga$0.6\,cm. This emission has a characteristic
bell-like spectral shape which can be easily distinguished from the power-law spectra of free-free and
synchrotron emission.

As discussed in Sect.~\ref{sect:recombination}, free-free thermal emission from photoevaporating disks is  optically thin at millimeter and centimeter wavelengths, meaning a radio spectral index $\alpha$ (for the gas only) of -0.1. 
Thus, the combination of sensitive EVLA continuum observations at 3.5\, and 6\,cm is the most straightforward way to constrain the radio slope of evolved disks and discriminate among the different mechanisms discussed here. However, the most direct way to test our inference of free-free emission from photoevaporating disks is via the detection of H recombination lines and associated blueshifts. 

\section{Hydrogen Radio Recombination Lines}\label{sect:recombination}
In a region that is fully or partially ionized electrons will be captured by protons to a state $n$ and undergo transitions to lower levels. We compute here the intensities of a representative sample of H$n\alpha$ recombination lines at millimeter and radio wavelengths as a function of $\Phi_{\rm EUV}$ and $L_{\rm X}$.

In our calculation, we consider spontaneous and internally stimulated emission (masing) but neglect  externally stimulated emission. The line flux density $F_l$ can be written as:
\begin{equation}\label{eq:sl}
F_l = B_{\nu}(T) \Omega_{HII} 
\left[ \frac{b_n \tau^{*}_{l} + \tau_c}{\tau_l + \tau_c} (1 - e^{-(\tau_l + \tau_c)}) - (1-e^{-\tau_c}) \right]
\end{equation}
where $B_{\nu}(T)$ is the Planck function at the gas temperature $T$, $\Omega_{HII}$ is the solid angle subtended by the ionized wind, $b_n$ is the LTE departure coefficient for the upper state $n$, and $\tau_c$, $\tau^{*}_l$, and $\tau_l$ are the continuum and line optical depths (the latter corrected for non-LTE effects) at the specific frequency $\nu_l$ \citep{bs78}. For the LTE departure coefficients we refer to \cite{sb79} for transitions $n \ge 50$ and to \cite{walmsley90} for lower $n$, i.e. for lines that fall at millimeter wavelengths. 

Eq.~\ref{eq:sl} simplifies greatly for the photoevaporative winds discussed in Sect.~\ref{sect:plausible} because we find line and continuum optical depths that are $<<1$ at millimeter and radio wavelengths. This also means that any masing is not significant because of the relatively low electron densities and pathlengths. To corroborate this statement let us consider the so-called turn-over wavelength $\lambda_T$, the wavelength at which $\tau_c=1$: $\lambda_T\sim 100/(T^{-1.35} n_e^2 2z_{HII})^{1/2}$ where all units are in cgs except for $z_{HII}$ which is in pc. One sees that even a partially ionized wind at 5,000\,K becomes optically thick at wavelengths $>20$\,cm for plausible wind values (Sect.~\ref{sect:plausible}). Hence, we can use the optically thin approximation to re-write eq.~\ref{eq:sl} as: $F_l = B_{\nu}(T) \Omega_{HII} b_n \tau^{*}_l$. The line optical depth can be written as $\tau^{*}_l = \tau_c \, r^{*}$, hence $F_l = F_c r^{*} b_n$ where $F_c$ is the thermal continuum flux.
Here, $r^{*}$ is the line-to-thermal continuum ratio in LTE {\bf at line center} assuming thermal broadening of the line:
\begin{equation}\label{eq:rstar}
r^{*}=2.33\times 10^{4} \Delta\nu_l^{-1} \nu_l^{2.1} T^{-1.15} E_l/E_c
\end{equation}
($\nu_l$ in GHz, $T$ in K) with $\Delta\nu_l$ being the line width\footnote{$\Delta\nu_l=2 \nu_l/c\sqrt{2ln(2) k T/m_h}$ with $c$ being the sound speed, $k$ the Boltzmann's constant, and $m_h$ the hydrogen mass} in kHz, and $E_l/E_c$ the ratio of line to continuum emission measure which we assume to be 0.9 following \cite{bs78}. 
Because $F_c$ is proportional to $\Phi_{\rm EUV}$ and $L_{\rm X}$ (Sect.~\ref{sect:plausible}) we obtain the following relations for the line flux densities assuming the Rayleigh Jeans approximation (which is valid for this hot gas even at millimeter wavelengths):
\begin{eqnarray}\label{eq:lineflux}
F_{l} = 2.1 \times 10^{-41} \, \nu_l b_n \, \left( \frac{51}{d} \right)^2\, \Phi_{\rm EUV} [\mu Jy]  \\
F_{l}=5.4\times10^{-31}\, \, \nu_l b_n \, \left( \frac{51}{d} \right)^2 \, L_{\rm X} \, \, \, [\mu Jy]
\end{eqnarray}
where $\nu_l$ is in GHz and $d$ is the distance in pc. 

\begin{figure}
\includegraphics[scale=0.8]{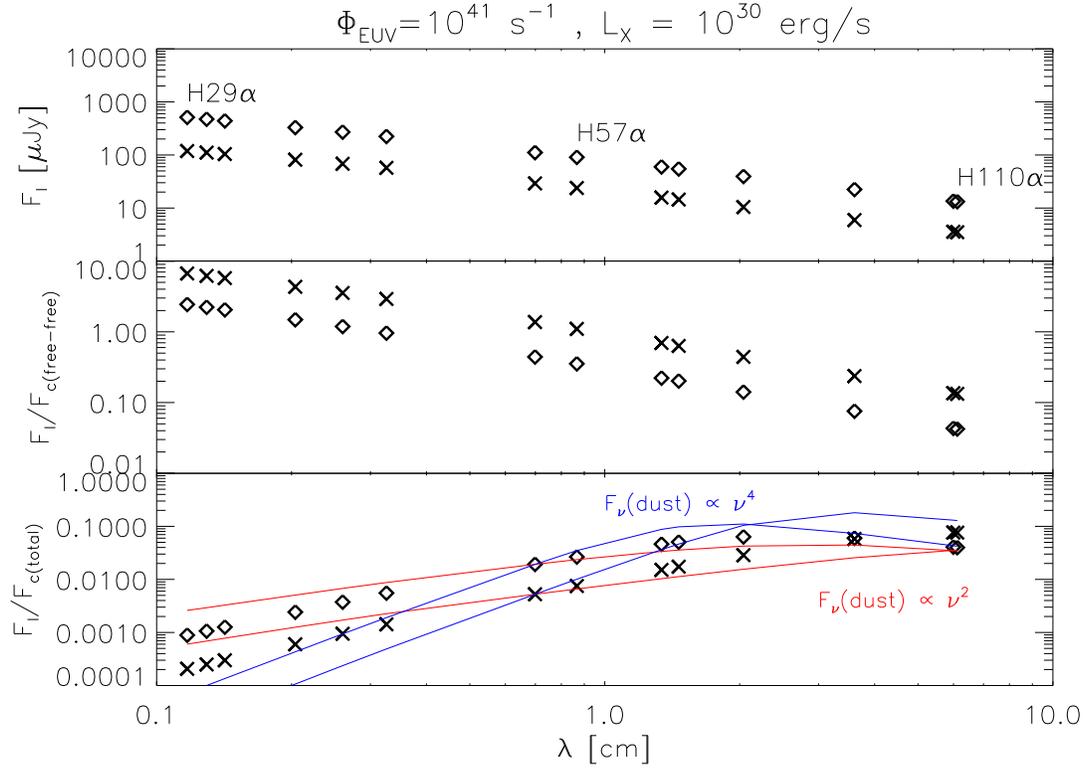}
\caption{First panel: Line fluxes for a representative sample of H$n\alpha$ recombination lines at centimeter and millimeter wavelengths. 
For the EUV-case (diamonds) we have assumed $\Phi_{EUV}=10^{41}$\,s$^{-1}$ while for the X-ray case (x) we have used $L_{\rm X}=10^{30}$\,erg/s.
Note that line fluxes are directly proportional to these quantities.
Second panel: Line {\bf center}-to-free-free continuum flux ratios for the same set of transitions. 
Third panel:  Line {\bf center}-to-total continuum flux ratios for the same set of transitions. For the assumed continuum emission we refer to Sect.~\ref{sect:recombination}. Symbols are for $\beta=0.6$, blue lines for  $\beta=2$, and red lines for $\beta=0$.
\label{fig:ratio}
}
\end{figure}

The upper panel of Fig.~\ref{fig:ratio} shows flux densities for a representative set of Hn$\alpha$ recombination lines for a fully ionized EUV-wind (10,000\,K gas) and a partially ionized X-ray wind (5,000\,K gas). We have taken $\Phi_{\rm EUV}=10^{41}$\,s$^{-1}$ and $L_{\rm X}=10^{30}$\,erg/s respectively and scaled the fluxes to  51\,pc. Line-to-free-free continuum ratios are shown in the panel below. They scale with the gas temperature as $\sim T^{-1.65}$ which explains why the cooler X-ray gas has higher line-to-free-free-continuum ratios. These ratios also increase as $\nu^{1.1} \times b_n$ (with $b_n$ slowly decreasing) as we move to high frequencies/short wavelengths suggesting that millimetric transitions might be the easiest to detect. However, the dust thermal emission increases more steeply at shorter wavelengths as illustrated in the lower panel of Fig.~\ref{fig:ratio} reducing the total line-to-continuum ratio. For the dust contribution we have taken here the mean 7\,mm flux of transitional disks in Taurus scaled at 51\,pc and then applied a power law of the form $F_\nu \propto \nu^{2+\beta}$ with $\beta=2,0.6,0$ to encompass the range of possible SED slopes (from an ISM-like to the average of Taurus disks to a SED dominated by large bodies with grey opacity). This plot illustrates that short cm wavelengths have actually the higher total line-to-continuum ratios. 

The EVLA could detect (but not spectrally resolve) some of these lines given the sensitivity of $\sim5 \,\mu$Jy rms in about 3 hours with a bandwidth of $\sim$500\,MHz. In the millimeter regime, we will be able to detect Hn$\alpha$ recombination lines only if the dust disk emission is spatially more extended than the free-free emission so that the line-to-continuum ratio can be increased at $r_g$, where the wind emission peaks. Evolved disks with dust gaps and large $\Phi_{\rm EUV}$ (or $L_{\rm X}$) are the best candidates to detect H recombination lines at millimeter wavelengths.

\section{Conclusions and Perspectives}\label{sect:conclusions}
This letter investigates the radio/millimeter properties of photoevaporative winds driven by high energy photons from the central star.
We show that free-free continuum emission from fully (EUV case) or partially ionized (X-ray case) winds is directly proportional to the stellar ionizing flux/Xray luminosity.
Given the inferred/measured $\Phi_{EUV}$/$L_{X}$ of young sun-like stars, centimeter wind emission should be detectable out to nearby star-forming regions. 
Other mechanisms producing centimeter emission can be ruled out with observations at multiple wavelengths measuring the radio spectral slope.
However, the smoking gun for free-free emission from photoevaporative winds would come from the detection and profiles of H radio recombination lines. Our calculations suggest that a few of them should be detectable at radio wavelengths and might be also detectable in the millimeter with ALMA if the dust contribution can be spatially resolved out using high resolution.

We point out that VLA observations might have already detected the free-free continuum emission from the nearby and photoevaporating disk of TW~Hya. Taking TW~Hya as a case study, we show how radio observations can be used to infer the ionizing luminosity reaching the disk, the wind temperature and electron density, and hence compute mass flow rates from star-driven photoevaporation. This latter parameter is essential to estimate disk lifetimes and understand whether photoevaporation is primarily driven by stellar X-rays at $10^{-8}$\,M$_\sun$/yr (e.g., \citealt{eo10}) or by EUV photons at about a hundred time lower rate (e.g., \citealt{alexander06}). Evolved disks with low mass accretion rates should be the prime targets to expand the analysis presented here and to identify what are the typical mass loss rates from star-driven photoevaporation. These empirically derived rates will enable evaluating the impact of photoevaporation on the dispersal of protoplanetary disks, as well as on planet formation and migration.

\acknowledgments
IP acknowledges support from the National Science Foundation through the research grant AST0908479.



{\it Facilities: \facility{VLA}, \facility{ALMA}}.



\clearpage



\clearpage

\clearpage







\clearpage

\clearpage




\end{document}